\documentclass[conference,twocolumn]{IEEEtran}

\def\paperversion{2} 

\IEEEoverridecommandlockouts

\usepackage{etex}

\usepackage[subtle,indent=normal,mathspacing=normal]{savetrees}

\usepackage{balance}
\usepackage[dvipsnames]{xcolor}

\usepackage[cmex10]{amsmath}
\usepackage{amssymb,amsfonts,amssymb,amsthm}
\usepackage{stmaryrd}
\usepackage{mathtools}

\makeatletter
\renewcommand*\env@matrix[1][*\c@MaxMatrixCols c]{%
  \hskip -\arraycolsep
  \let\@ifnextchar\new@ifnextchar
  \array{#1}}
\makeatother
\usepackage[T1]{fontenc}
\usepackage{cite}
\usepackage{algorithmic}
\usepackage{graphicx}
\usepackage{textcomp}
\usepackage{algorithm2e}
\RestyleAlgo{ruled}
\SetKwComment{Comment}{/* }{ */}
\SetKwInput{KwInput}{Input}
\SetKwInput{KwOutput}{Output}

\usepackage{url}
\usepackage{enumerate}
\usepackage{float}
\usepackage{booktabs}
\usepackage{multirow}
\usepackage{colortbl}
\usepackage[font=footnotesize]{caption}
\usepackage[font=footnotesize]{subcaption}

\usepackage{resizegather}

\usepackage{cleveref}
\Crefname{figure}{Fig.}{Figs.}

\usepackage{pgfplots}
\usepackage{pgfplotstable}
\usepackage{tikz}
\usetikzlibrary{calc,positioning,decorations.pathreplacing}

\DeclareCaptionLabelSeparator{periodspace}{.\quad}
\def\BibTeX{{\rm B\kern-.05em{\sc i\kern-.025em b}\kern-.08em
    T\kern-.1667em\lower.7ex\hbox{E}\kern-.125emX}}

\makeatletter
\newcommand{\removelatexerror}{\let\@latex@error\@gobble}
\makeatother

\DeclareMathOperator{\bin}{Bin}
\DeclareMathOperator{\lcm}{lcm}

\newcommand{\pmac}{P_{\text{DL}}}
\newcommand{\plr}{\mathsf{PLR}}

\begin{document}

\title{Packet Aggregation May Harm Batched Network Coding}
\author{\IEEEauthorblockN{Hoover~H.~F.~Yin}
\thanks{H.~H.~F.~Yin is with the Department of Information Engineering, The Chinese University of Hong Kong.}
}

\maketitle

\begin{abstract}
Batched network coding (BNC) is a solution to multi-hop transmission on networks with packet loss.
To be compatible with the existing infrastructure, BNC is usually implemented over UDP.
A single error bit will probably result in discarding the packet.
UDP-Lite is a variant of UDP that supports partial checksums.
As long as the data covered by the checksum is correct, damaged payload will be delivered.
With UDP-Lite, we can cope with other techniques such as payload aggregation of BNC packets to reduce the protocol overhead, and forward error correction to combat against bit errors.
Unlike traditional transmissions, BNC has a loss resilience feature and there are dependencies between BNC packets.
In this paper, we conduct a preliminary investigation on BNC over UDP-Lite.
We show that aggregating as much as we can is not always the best strategy, and a hop-by-hop distributed efficiency optimization approach may lead to a worse throughput compared with the scheme without aggregation in a long network.
These unnatural results caution that a casual integration of techniques with BNC can be harmful, and give us hints on future research directions.
\end{abstract}

\section{Introduction}

In computer networking, application data is encapsulated by a few layers of protocols in a network stack, with one or more layers providing integrity verification.
When bit errors are detected, the entire packet is dropped by default.
A packet sent by a source node can arrive at a sink node if it is not lost at any of the network links.
If the packet loss rate per link is significant, the chance of a packet arriving at the sink node decreases as the number of hops increases.
To achieve reliable communications, end-to-end retransmission is a standard approach. 
However, it is 
not efficient when the overall packet loss rate is significant.

\emph{Batched network coding (BNC)} \cite{mahdaviani12,tang18,yang14bats} is a class of linear network coding \cite{flow,linear} aiming for practical realization on multi-hop networks with packet loss.
BNC encodes the whole piece of data to be sent into multiple batches, and then restricts the application of random linear network coding (RLNC) \cite{random2} to the packets belonging to the same batch.
Instead of forwarding, recoding is performed at the intermediate nodes to generate more packets.
There exists BNC that can achieve a close-to-optimal rate, e.g., BATS codes \cite{yang14bats}.

When deploying BNC, a common choice is to implement it over existing transport layer or application layer protocols.
Besides being compatible with the existing network infrastructure, the lower-layer problems such as routing can be separated.
The headers of these protocols, e.g., UDP and IP, induce extra overhead to the transmission.
To reduce the protocol overhead, aggregation is an efficient approach \cite{sze02aggregate,deng13aggregate,majeed12aggregate}.
That is, we can aggregate a few BNC packets into a single payload for the lower-layer protocol. 
Further, to prevent discarding all aggregated packets due to a single-bit error, we can use protocols such as UDP-Lite \cite{stanislaus05udp,shukla2016udp,singh01udp} that allow passing potentially damaged data payload to the BNC recoder/decoder, with a cost of extra integrity check or error correction per BNC packet.

With BNC encoding/recoding, there are dependencies between BNC packets within the same batch.
The next node only needs to receive a certain number of BNC packets per batch to retrieve all ``information'' carried by this batch.
Due to the variation of ``information'' carried by each batch and the chance of losing all aggregated packets due to corruption of protocol headers, the number of aggregated BNC packets affects the expected number of ``useful'' BNC packets arriving at the next node.
In this paper, we investigate:
\begin{enumerate}
    \item How should we model the efficiency of aggregated BNC packets, i.e., ``useful'' data size divided by transmitted size?
    \item How many BNC packets are to be aggregated to maximize the efficiency?
    \item What is the effect to the throughput when aggregating BNC packets?
    \item Can error correction code help in terms of throughput?
\end{enumerate}

Our results show that aggregating as much as we can is not always the best strategy.
The efficiency at each hop tends to increase when aggregating more BNC packets, but there are some troughs in the function.
When optimizing the efficiency at each hop distributively using local batch statistics, the throughput becomes worse than the one without aggregation after a few hops.
These results suggest that integrating existing techniques casually with BNC can be harmful.

\section{System Model}

Although there may be multiple paths connecting the source node and the sink node, one path will be chosen after routing in a common unicast setting.
Therefore, we consider a line network in this paper, which is formed by a chain of network nodes. 

The packet flow is not conserved as each node performs recoding on the received BNC packets to generate new (or more) BNC packets. 
The aggregated BNC packets are disaggregated for recoding.
That is, each node has the freedom to choose a different number of BNC packets to be aggregated to optimize the efficiency.

\subsection{Batched Network Coding (BNC)}

The data to be transmitted is divided into a set of \emph{input packets}, each of which is regarded as a vector of $K$ symbols over a fixed finite field.
To minimize the overhead like zero-padding overhead, we can apply certain optimization designed for BNC \cite{yin22size}.
We consider a sufficiently large field size so that we can assume two random vectors over this field are linearly independent of each other with high probability.
This assumption is widely adopted in network coding literature like \cite{zhou17b,yin22mp,coupon,field_size} to simplify the analysis.
With this assumption, any $n$ random vectors in an $r$-dimensional vector space can span a $\min\{n, r\}$-dimensional vector space with high probability.
Practically, $GF(256)$ is sufficiently large for BNC \cite{bar}.

A BNC encoder is run at the source node to generate batches from the input packets.
Each batch consists of $M$ coded packets, where this integer $M > 0$ is also known as the \emph{batch size}.
To generate a batch, a subset of input packets is chosen.
The way to select the subset depends on the application and code design.
Each coded packet in a batch is the random linear combinations on the chosen subset of input packets.
A coefficient vector is also attached to each packet.
The purpose of the coefficient vectors is to record the recoding operations at the intermediate nodes.
Two coded packets in a batch are said to be linearly independent of each other if and only if their coefficient vectors are linearly independent of each other.
Right after a batch is generated, the coded packets within it are assigned as linearly independent of each other by initializing the coefficient vectors properly \cite{yang22pro,yin20pro}.

\begin{figure}
    \centering
    \scriptsize
    \begin{tikzpicture}[minimum height=4mm]
        \node[draw,minimum width=15mm] (id) {batch ID};
        \node[draw,minimum width=15mm,right=0cm of id] (co) {coeff. vector};
        \node[draw,minimum width=50mm,right=0cm of co] (pl) {payload (coded packet)};
        \draw [decorate,decoration={brace,amplitude=3pt,mirror}] ($(id.south west)+(0,-.05)$) -- ($(co.south east)+(0,-.05)$) node [black,midway,below] {BNC header};
        \draw [decorate,decoration={brace,amplitude=3pt}] ($(id.north west)+(0,.05)$) -- ($(pl.north east)+(0,.05)$) node [black,midway,above] {BNC packet};
    \end{tikzpicture}
    \vskip -1em
    \caption{An example of BNC header and BNC packet. The number of data fields in a BNC header varies from protocol to protocol.}
    \label{fig:bnc_packet}
\end{figure}
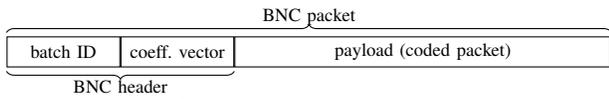

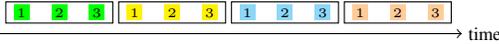
\begin{figure}
    \centering
    \scriptsize
    \begin{tikzpicture} 
        \foreach \j in {1,...,3}
        {
            \fill[green] (0.5*\j-0.5,0) rectangle node[name=a\j,color=black] {{\tiny $\j$}} ++(6pt,6pt);
        }
        \draw ($(a1)+(-6pt,4pt)$) rectangle node {} ($(a3)+(6pt,-4pt)$);
        \foreach \j in {1,...,3}
        {
            \fill[yellow] (1.5+0.5*\j-0.5,0) rectangle node[name=b\j,color=black] {{\tiny $\j$}} ++(6pt,6pt);
        }
        \draw ($(b1)+(-6pt,4pt)$) rectangle node {} ($(b3)+(6pt,-4pt)$);
        \foreach \j in {1,...,3}
        {
            \fill[cyan!40] (3+0.5*\j-0.5,0) rectangle node[name=c\j,color=black] {{\tiny $\j$}} ++(6pt,6pt);
        }
        \draw ($(c1)+(-6pt,4pt)$) rectangle node {} ($(c3)+(6pt,-4pt)$);
        \foreach \j in {1,...,3}
        {
            \fill[orange!40] (4.5+0.5*\j-0.5,0) rectangle node[name=d\j,color=black] {{\tiny $\j$}} ++(6pt,6pt);
        }
        \draw ($(d1)+(-6pt,4pt)$) rectangle node {} ($(d3)+(6pt,-4pt)$);
        \draw[->] ($(a1)+(-10pt,-8pt)$) -- ($(d3)+(10pt,-8pt)$) node[right] {time};
    \end{tikzpicture}
    \vskip -1em
    \caption{An illustration of sending batches sequentially. Each rectangle is a batch, and each square is a BNC packet.}
    \label{fig:scheduling}
\end{figure}

Besides the coded packets (payload) and the coefficient vectors, the BNC decoder at the sink node requires some more information for decoding.
For example, the minimal BNC protocol header also consists of a batch ID to distinguish different batches \cite{yang14a,yin19pro2}.
Sophisticated BNC protocols include more data fields such as coding parameters and data stream identification \cite{fun,rfc_bats,yin22rec}.
We call the above data fields excluding the payload the \emph{BNC header}.
A BNC header together with the payload is called a \emph{BNC packet}.
An example based on the minimal BNC protocol is illustrated in \Cref{fig:bnc_packet}.
As a preliminary investigation, we use a simple scheduling strategy that the batches are sent sequentially.
This is depicted in \Cref{fig:scheduling}.

The number of linearly independent packets in a batch, known as the \emph{rank} of the batch, is a measure of information carried by the batch.
In an optimized BNC, e.g., BATS code, each ``rank'' almost corresponds to the recovery of a distinct input packet.
Therefore, the performance of BNC is usually measured by the expected rank among all batches arriving at the sink node.

At the source node, a freshly generated batch has rank $M$.
Due to packet loss, the rank of a batch monotonically decreases through the network.
To slow down the dropping rate of rank, \emph{recoding} is performed at the intermediate nodes.
After receiving all packets of a batch that are not dropped, recoded packets of this batch are generated by taking random linear combinations on the received packets of this batch.
The number of recoded packets per batch can be different, depending on the extra information known at the intermediate node \cite{adaptive,scheduling,ge_adaptive,uni,bar}.
For the simplicity of analysis and implementation, many recoding schemes generate $M$ recoded packets per batch \cite{fun,zhou17b,bats_schedule}.

At the sink node, a BNC decoder is run.
The decoding method depends on the way to form batches.
Common algorithms include Gaussian elimination, belief propagation, inactivation, etc.

\subsection{Frame Structure}

We consider that the BNC packets are encapsulated by UDP-Lite.
As we do not want the datagram to be dropped due to the corruption of any of the encapsulated BNC packets, we configure that only the UDP-Lite header is checksummed.
To ensure the integrity of each BNC packet that is not covered by the checksum of UDP-Lite, we can either attach a checksum or use an error correction code.

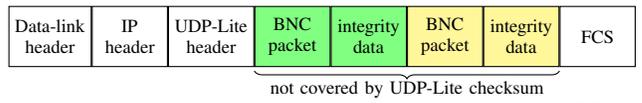
\begin{figure}
    \centering
    \scriptsize
    \begin{tikzpicture}[minimum height=8mm,every text node part/.style={align=center}]
        \node[draw,minimum width=10mm] (dl) {Data-link\\header};
        \node[draw,minimum width=10mm,right=0cm of dl] (ip) {IP\\header};
        \node[draw,minimum width=10mm,right=0cm of ip] (h0) {UDP-Lite\\header};
        \foreach \j/\c[evaluate=\j as \k using int(\j-1)] in {1/green!50,2/yellow!50}
        {
            \node[draw,minimum width=10mm,right=0cm of h\k,fill=\c] (b\j) {BNC\\packet};
            \node[draw,minimum width=10mm,right=0cm of b\j,fill=\c] (h\j) {integrity\\data};
        }
        \node[draw,minimum width=10mm,right=0cm of h2] (fcs) {FCS};
        \draw [decorate,decoration={brace,amplitude=3pt,mirror}] ($(b1.south west)+(0,-.05)$) -- ($(h2.south east)+(0,-.05)$) node [black,midway,below=1pt,minimum height=4mm] {not covered by UDP-Lite checksum};
    \end{tikzpicture}
    \vskip -1em
    \caption{A frame with FCS and $2$ aggregated BNC packets over UDP-Lite.}
    \label{fig:whole_packet}
\end{figure}

The data-link layer protocol depends on the link connection.
The simplest case is IEEE 802.3 (Ethernet).
By default, the frame check sequence (FCS) is used to detect errors for dropping corrupted frames.
In other words, UDP-Lite requires an operating system that supports passing frames with failed FCS or supports partial FCS schemes.
An example of an Ethernet frame carrying $2$ aggregated BNC packets over UDP-Lite is shown in \Cref{fig:whole_packet}.

\subsection{Notations}

\begin{table}
    \let\oldtabcolsep\tabcolsep
    \setlength{\tabcolsep}{4pt}
    \centering
    \caption{Table of Notations}
    \label{tab:notations}
    \vskip -0.9em
    \begin{tabular}{cl|cl}
        \toprule
        $M$ & batch size & $N$ & no. of aggregated BNC packets\\
        $H$ & BNC header size & $L$ & max. frame payload size\\ 
        $K$ & BNC payload size & $P$ & IP plus UDP/UDP-Lite header size\\
        $S(N)$ & frame size & $\pmac$ & frame size that can't be corrupted\\
        $F$ & integrity data size & $Q$ & frame size that can be corrupted\\
        $p$ & bit error rate & $d$ & prob. of recv. non-corruptable parts\\
        $\hbar_r$ & p.m.f. of rank dist. 
        & $f$ & prob. of recv. a BNC packet\\
    \end{tabular}
    \setlength{\tabcolsep}{\oldtabcolsep}
\end{table}

The frequently used notations in the main text are summarized in \Cref{tab:notations}. 
Denote the set of positive integers by $\mathbb{Z}^+$.
Let $L$ be the maximum size of a network layer packet.
The value depends on the data-link layer protocol and the support of jumbo frame.
For example, $L = 9000$ and $L = 1500$ are the values of maximum transmission unit (MTU) of IEEE 802.3 with and without the support of jumbo frame respectively.
Let $P$ be the total number of bytes in the network layer and transport layer headers.
These bytes are checksummed and must be received correctly.
The value depends on the protocol in use.
For example, $P = 28$ for UDP-Lite over IPv4 or $P = 48$ for UDP-Lite over IPv6.
Other combinations include robust header compression (ROHC) for UDP-Lite (RFC 4019). 

The FCS in data-link frames might be ignored when using partial checksum protocols.
Let $\pmac$ be the number of bytes in the data-link layer per frame that must be received correctly, and let $Q$ be the number of bytes in the data-link layer per frame that can be corrupted or ignored.
For example, without counting the FCS, we have $\pmac = 22$ for IEEE 802.3, which includes the preamble and the start frame delimiter in the Layer 1 and Layer 2 Ethernet frame headers without using 802.1Q tag.
The corresponding $Q = 4$ is the size of the FCS.

Let $H$ be the number of bytes of the header for each BNC packet.
For example, $H = 2+M$ in a minimal BNC protocol with a $2$-bytes batch ID and a length-$M$ coefficient vector over $GF(256)$. 
Let $F$ be the number of bytes for integrity check and/or error correction per BNC packet.
In other words, when the input packet size is $K$, the length of a BNC packet is $H+K+F$ bytes.

Let $N$ be the number of BNC packets aggregated in one frame.
Let $S(N)$ be the number of bytes of a frame aggregated with $N$ BNC packets, i.e.,
    $S(N) = P + \pmac + Q + N(H+K+F)$.

Next, for simplicity, assume each bit has an error rate $p$ independently.
Most BNC literature considered packet loss rate (PLR) instead of bit error rate (BER) because in these works, each packet has the same length and consists of $1$ BNC packet only, i.e., it is equivalent to consider PLR or BER.
We call this PLR the \emph{baseline PLR}, which can be formulated as $\plr = 1 - (1-p)^{8(P+\pmac+Q+H+K)}$.
As the aggregated packet has a different length from that without aggregation, it has a different loss rate.
We convert the baseline PLR to BER by
    $p = 1 - (1-\plr)^{\frac{1}{8(P+\pmac+Q+H+K)}}$.
Based on the BER, let
    $d = (1-p)^{8(P+\pmac)}$
be the probability of correctly receiving the protocol headers that cannot be corrupted. 

Denote by
    $\bin(k; n, x) = \binom{n}{k} x^k (1-x)^{n-k}$
the probability mass of a binomial distribution. 
Denote by
\begin{equation*}
    \bin_d(k; n, f) = \begin{cases}
        d\binom{n}{k} f^k (1-f)^{n-k} & \text{if } 0 < k \le n,\\
        (1-d) + d (1-f)^n & \text{if } k = 0
    \end{cases}
\end{equation*}
the probability mass of successfully receiving $k$ out of $n$ BNC packets of the same batch packed in another UDP-Lite packet, where $f$ is the probability of receiving a BNC packet correctly.
The formulation of $f$ depends on the error correction ability.
Say, if there is no error correction code applied, then
    $f = (1-p)^{8(H+K+F)}$.
If a $\lfloor F/2 \rfloor$-error correcting code with $F$-byte redundancy over $GF(256)$ (as an example) is applied to each BNC packet independently, then we have
    $f = \sum_{e = 0}^{\lfloor F/2 \rfloor} \bin(e; H+K+F, 1-(1-p)^8)$.

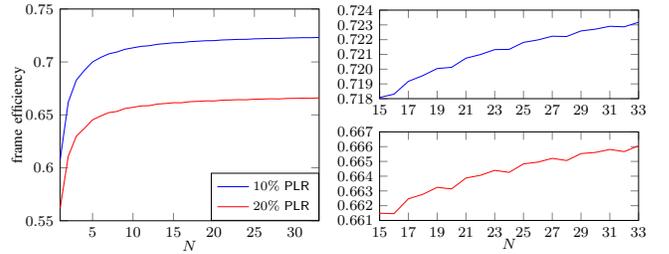
\begin{figure}
\centering
\begin{minipage}{.225\textwidth}
\centering
\begin{tikzpicture}[scale=.7]
    \begin{axis}[
        small,
        xtick={5,10,...,35},
        ytick={0.55,0.6,...,0.75},
        xmin=1, xmax=33,
        ymin=0.55, ymax=0.75,
        xlabel=$N$,
        ylabel=frame efficiency,
        x label style={at={(axis description cs:.5,.07)},anchor=north},
        y label style={at={(axis description cs:.14,.5)},anchor=south},
        label style={font=\footnotesize},
        legend style={legend columns=1,font=\scriptsize,anchor=south east,at={(1,0)}}
    ]
        \addplot[blue] table [x=N, y=a1, col sep=comma] {data/frame.csv};
        \addlegendentry{$10\%$ $\plr$}
        \addplot[red] table [x=N, y=a2, col sep=comma] {data/frame.csv};
        \addlegendentry{$20\%$ $\plr$}
    \end{axis}
\end{tikzpicture}
\end{minipage}~
\begin{minipage}{.225\textwidth}
\centering
\begin{tikzpicture}[scale=.7]
    \begin{axis}[
        small,
        height=0.8\textwidth,
        xtick={15,17,...,35},
        ytick={0.718,0.719,...,0.724},
        xmin=15, xmax=33,
        ymin=0.718, ymax=0.724,
        yticklabel style={/pgf/number format/.cd,fixed,fixed zerofill,precision=3,/tikz/.cd},
        label style={font=\footnotesize},
        legend style={legend columns=2,font=\scriptsize}
    ]
        \addplot[blue] table [x=N, y=a1, col sep=comma] {data/frame.csv};
    \end{axis}
\end{tikzpicture}\\
\begin{tikzpicture}[scale=.7]
    \begin{axis}[
        small,
        height=0.8\textwidth,
        xtick={15,17,...,35},
        ytick={0.661,0.662,...,0.667},
        xmin=15, xmax=33,
        ymin=0.661, ymax=0.667,
        xlabel=$N$,
        x label style={at={(axis description cs:.5,.18)},anchor=north},
        yticklabel style={/pgf/number format/.cd,fixed,fixed zerofill,precision=3,/tikz/.cd},
        label style={font=\footnotesize},
        legend style={legend columns=2,font=\scriptsize}
    ]
        \addplot[red] table [x=N, y=a2, col sep=comma] {data/frame.csv};
    \end{axis}
\end{tikzpicture}
\end{minipage}
    \vskip -0.7em
\caption{Frame efficiency when $\plr = 10\%$ and $20\%$ respectively, with $M = 4$ and $K = 256$. Note that the frame efficiency is not monotonically increasing.}
\label{fig:frame1}
\end{figure}

\section{Frame Efficiency}

At any non-sink node, if a packet arriving at the next node is not redundant, the payload data excluding all the headers is called ``useful''.
The efficiency of a frame is defined as the expected ``useful'' data size divided by the transmitted size.
That is, the loss of data is considered in this definition.
This way, the frame efficiency is equivalent to the throughput per byte sent.

Fix a non-sink node.
Our objective is to find the number of BNC packets aggregated in one frame, $N$, that can maximize the efficiency.
As the BNC packets of a batch can be separated into multiple UDP-Lite packets, some ``ranks'' of this batch may be received by the next node already.
This way, we need to consider the rank increment the batch in a UDP-Lite packet can provide the next node.
By only considering the BNC packets packed in a UDP-Lite packet, let $E$ be the sum of expected rank increment of these batches at the next node.
That is, $E$ measures the amount of information received by the next node.
The efficiency of a frame can be formulated as
\begin{equation*}
    \max_{N \in \mathbb{Z}^+} \frac{dKE}{S(N)} \text{ s.t. } P+N(H+K+F) \le L,
\end{equation*}
where $dKE$ is the expected useful data size.
\ifnum\paperversion=1
The analytical formulation of $E$ can be found in the appendix of \cite{udp_full}. 
\fi
\ifnum\paperversion=2
The analytical formulation of $E$ can be found in \Cref{sec:analysis}.
\fi

\subsection{Optimal Aggregation}

For demonstration purposes, we use jumbo Ethernet frames, UDP-Lite over IPv4 without ROHC, 
$GF(256)$ as the finite field for BNC, and apply the minimal BNC protocol, i.e., 
$L = 9000$, $P = 28$, $\pmac = 22$, $Q = 4$ and $H = M+2$.

In this subsection, we give an overview of frame efficiency for different $N$.
We do not consider the use of error correction code here, but we apply the Internet checksum (i.e., $F = 2$) to verify the integrity of BNC packet.
Let $(\hbar_r)_{r = 1}^M$ be the rank distribution of the batches arriving at the current node.
For demonstration purposes, let $\hbar_r = \frac{\bin(r; M, 0.8)}{\sum_{i = 1}^M \bin(i; M, 0.8)}$.
Our first example considers $M = 4$ and $K = 256$.
The valid range of $N$ is $1 \le N \le 33$.
\Cref{fig:frame1} shows the frame efficiency when $\plr = 10\%$ and $20\%$.
We can see that although the trend is increasing when $N$ increases, the efficiency is not an increasing function.

\begin{figure}
\centering
\begin{minipage}{.225\textwidth}
\centering
\begin{tikzpicture}[scale=.7]
    \begin{axis}[
        small,
        xtick={10,20,...,80},
        ytick={0.41,0.43,...,0.61},
        xmin=1, xmax=76,
        ymin=0.41, ymax=0.62,
        xlabel=$N$,
        ylabel=frame efficiency,
        x label style={at={(axis description cs:.5,.07)},anchor=north},
        y label style={at={(axis description cs:.1,.5)},anchor=south},
        label style={font=\footnotesize},
        legend style={legend columns=2,font=\scriptsize}
    ]
        \addplot[black] table [x=N, y=a, col sep=comma] {data/frame2.csv};
    \end{axis}
\end{tikzpicture}
\end{minipage}~
\begin{minipage}{.225\textwidth}
\centering
\begin{tikzpicture}[scale=.7]
    \begin{axis}[
        small,
        xtick={40,44,...,76},
        ytick={0.601,0.602,...,0.605},
        xmin=40, xmax=76,
        ymin=0.601, ymax=0.6055,
        xlabel=$N$,
        x label style={at={(axis description cs:.5,.07)},anchor=north},
        yticklabel style={/pgf/number format/.cd,fixed,fixed zerofill,precision=3,/tikz/.cd},
        label style={font=\footnotesize},
        legend style={legend columns=2,font=\scriptsize}
    ]
        \addplot[black] table [x=N, y=a, col sep=comma] {data/frame2.csv};
    \end{axis}
\end{tikzpicture}
\end{minipage}
    \vskip -0.7em
\caption{Frame efficiency when $\plr = 25\%$, $M = 4$ and $K = 110$. A sudden drop can be observed at $N = 76$, which makes the optimal $N = 75$.}
\label{fig:frame2}
\end{figure}
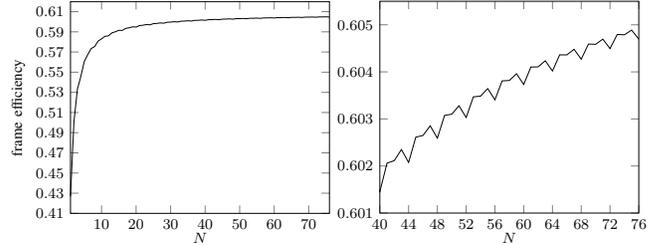

The zig-zag shape is due to the linear dependency of BNC packets across UDP-Lite packets, i.e., the shape is mainly due to $E$.
The consequence is that the largest $N$ does not always give the highest efficiency.
For example, consider $M = 4$, $K = 110$ and $\plr = 25\%$.
The valid range of $N$ is $1 \le N \le 76$, while the efficiency is maximized when $N = 75$.
\Cref{fig:frame2} shows the frame efficiency.
We can clearly see that there is a drop at $N = 76$.

The above examples suggest that in theory, it is not always the best for BNC to aggregate as much as we can.
However, the drop in efficiency is not significant.
That is, adopting the largest possible $N$ is a good heuristic for frame efficiency.
This way, we do not need to know the rank distribution in practice.

\subsection{Throughput}

The most concerned performance measure is the throughput per byte sent, i.e., the frame efficiency.
In this subsection, we compare the frame efficiency with and without using aggregation.
When there is no aggregation, we can still apply an error correction code to the only BNC packet in a UDP-Lite packet.
Therefore, we also compare the cases with and without error correction.
More specifically, we use the Internet checksum ($F = 2$) when there is no error correction, or otherwise use a $1$-error correcting code with $3$-byte redundancy over $GF(256)$, i.e., $F = 3$, for each BNC packet.
At first glance, the frame efficiency should be boosted when we aggregate some BNC packets.
However, it is not always the case.

We consider a $10$-hop line network where each link has the same PLR.
In the following discussion, each node decides the value of $N$ individually using its local batch statistics.
We reuse the parameters 
$L = 9000$, $P = 28$, $\pmac = 22$, $Q = 4$ and $H = M+2$.

\begin{figure}
\centering
\begin{minipage}{.225\textwidth}
\centering
\begin{tikzpicture}[scale=.7]
    \begin{axis}[
        small,
        xtick={1,2,...,10},
        ytick={0.3,0.35,...,1},
        xmin=1, xmax=10,
        ymin=0.3, ymax=1,
        xlabel=hops,
        ylabel=throughput / byte sent,
        x label style={at={(axis description cs:.5,.07)},anchor=north},
        y label style={at={(axis description cs:.12,.5)},anchor=south},
        label style={font=\footnotesize},
        legend style={legend columns=2,font=\tiny,at={(0,0)},anchor=south west}
    ]
        \node[anchor=north east] at (axis description cs:1,1) {$\plr = 10\%$};
        \addplot[blue] table [x=N, y=a1, col sep=comma] {data/tp.csv};
        \addlegendentry{optimal $N$, checksum}
        \addplot[only marks,blue,mark=o] table [x=N, y=a2, col sep=comma] {data/tp.csv};
        \addlegendentry{largest $N$, checksum}
        \addplot[red] table [x=N, y=a3, col sep=comma] {data/tp.csv};
        \addlegendentry{$N = 1$, checksum}
        \addplot[blue,dashed] table [x=N, y=b1, col sep=comma] {data/tp.csv};
        \addlegendentry{optimal $N$, FEC}
        \addplot[only marks,blue,mark=triangle] table [x=N, y=b2, col sep=comma] {data/tp.csv};
        \addlegendentry{largest $N$, FEC}
        \addplot[red,dashed] table [x=N, y=b3, col sep=comma] {data/tp.csv};
        \addlegendentry{$N = 1$, FEC}
    \end{axis}
\end{tikzpicture}
\end{minipage}~
\begin{minipage}{.225\textwidth}
\centering
\begin{tikzpicture}[scale=.7]
    \begin{axis}[
        small,
        xtick={1,2,...,10},
        ytick={0.3,0.35,...,1},
        xmin=1, xmax=10,
        ymin=0.3, ymax=1,
        xlabel=hops,
        x label style={at={(axis description cs:.5,.07)},anchor=north},
        label style={font=\footnotesize},
    ]
        \node[anchor=north east] at (axis description cs:1,1) {$\plr = 20\%$};
        \addplot[blue] table [x=N, y=c1, col sep=comma] {data/tp.csv};
        \addplot[only marks,blue,mark=o] table [x=N, y=c2, col sep=comma] {data/tp.csv};
        \addplot[red] table [x=N, y=c3, col sep=comma] {data/tp.csv};
        \addplot[blue,dashed] table [x=N, y=d1, col sep=comma] {data/tp.csv};
        \addplot[only marks,blue,mark=triangle] table [x=N, y=d2, col sep=comma] {data/tp.csv};
        \addplot[red,dashed] table [x=N, y=d3, col sep=comma] {data/tp.csv};
    \end{axis}
\end{tikzpicture}
\end{minipage}
\vskip -0.7em
\caption{The throughput of different $\plr$, $N$ and $F$, when $M = 4$ and $K = 256$.}
\label{fig:tp1}
\end{figure}
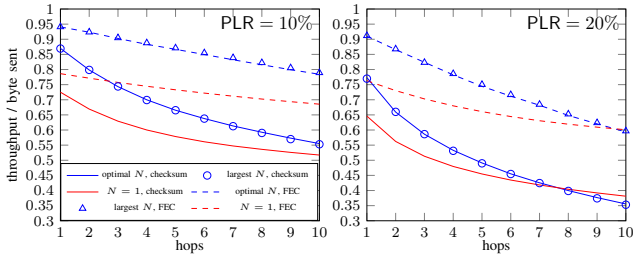

At the source node, all batches have rank $M$.
The rank distribution at each node is due to the packet loss events at the links.
As our first example, consider
$M = 4$ and $K = 256$.
The throughput is plotted in \Cref{fig:tp1} for $\plr = 10\%$ and $20\%$.
We can see that at the beginning, the frame efficiency with aggregation is higher than that without aggregation ($N = 1$).
However, the drop rate of the efficiency across the network is also greater.
When the PLR is higher, we can see that after certain number of hops, the frame efficiency with aggregation becomes lower than that without aggregation.
When error correcting code, the throughput is always significantly better compared with that without the code.
This is as expected because the loss rate is decreased.
However, the same issue about the drop rate of efficiency persists.

\begin{figure}
\centering
\begin{minipage}{.225\textwidth}
\centering
\begin{tikzpicture}[scale=.7]
    \begin{axis}[
        small,
        xtick={1,2,...,10},
        ytick={0.3,0.35,...,1},
        xmin=1, xmax=10,
        ymin=0.3, ymax=1,
        xlabel=hops,
        ylabel=throughput / byte sent,
        x label style={at={(axis description cs:.5,.07)},anchor=north},
        y label style={at={(axis description cs:.12,.5)},anchor=south},
        label style={font=\footnotesize},
        legend style={legend columns=2,font=\tiny,at={(0,0)},anchor=south west}
    ]
        \node[anchor=north east] at (axis description cs:1,1) {$M = 16$};
        \addplot[blue] table [x=N, y=a1, col sep=comma] {data/tp2.csv};
        \addlegendentry{optimal $N$, checksum}
        \addplot[only marks,blue,mark=o] table [x=N, y=a2, col sep=comma] {data/tp2.csv};
        \addlegendentry{largest $N$, checksum}
        \addplot[red] table [x=N, y=a3, col sep=comma] {data/tp2.csv};
        \addlegendentry{$N = 1$, checksum}
        \addplot[blue,dashed] table [x=N, y=b1, col sep=comma] {data/tp2.csv};
        \addlegendentry{optimal $N$, FEC}
        \addplot[only marks,blue,mark=triangle] table [x=N, y=b2, col sep=comma] {data/tp2.csv};
        \addlegendentry{largest $N$, FEC}
        \addplot[red,dashed] table [x=N, y=b3, col sep=comma] {data/tp2.csv};
        \addlegendentry{$N = 1$, FEC}
    \end{axis}
\end{tikzpicture}
\end{minipage}~
\begin{minipage}{.225\textwidth}
\centering
\begin{tikzpicture}[scale=.7]
    \begin{axis}[
        small,
        xtick={1,2,...,10},
        ytick={0.3,0.35,...,1},
        xmin=1, xmax=10,
        ymin=0.3, ymax=1,
        xlabel=hops,
        x label style={at={(axis description cs:.5,.07)},anchor=north},
        label style={font=\footnotesize},
    ]
        \node[anchor=north east] at (axis description cs:1,1) {$M = 8$};
        \addplot[blue] table [x=N, y=c1, col sep=comma] {data/tp2.csv};
        \addplot[only marks,blue,mark=o] table [x=N, y=c2, col sep=comma] {data/tp2.csv};
        \addplot[red] table [x=N, y=c3, col sep=comma] {data/tp2.csv};
        \addplot[blue,dashed] table [x=N, y=d1, col sep=comma] {data/tp2.csv};
        \addplot[only marks,blue,mark=triangle] table [x=N, y=d2, col sep=comma] {data/tp2.csv};
        \addplot[red,dashed] table [x=N, y=d3, col sep=comma] {data/tp2.csv};
    \end{axis}
\end{tikzpicture}
\end{minipage}
\vskip -0.7em
\caption{The throughput of different $M$, $N$ and $F$, when $\plr = 20\%$ and $K = 256$.}
\label{fig:tp2}
\end{figure}
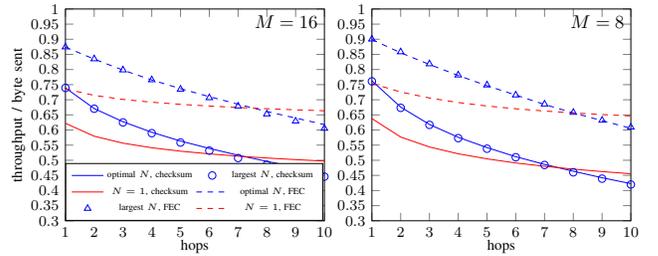

Although $M = 4$ is a possible choice in practice to reduce the latency, the throughput is better with a larger $M$.
It is also common to use $M = 8$ and $M = 16$.
\Cref{fig:tp2} shows the throughput when $\plr = 20\%$ and $K = 256$.
We can observe a similar conclusion in the plots.

One explanation of the higher drop rate of frame efficiency with aggregation is that, when losing all BNC packets in a UDP-Lite packet due to the corruption of headers, the rank of some batches drop significantly.
Although BNC aims to retain the rank of the batch, this effect is an irreversible damage and it propagates through the network, thus further decreases the throughput.

\section{Concluding Remarks}

By using partial-checksum feature of UDP-Lite, we can aggregate multiple BNC packets in a single UDP-Lite packet and/or apply error correction codes to recover more BNC packets.
However, we showed that a casual integration of techniques with BNC can be harmful.

\ifnum\paperversion=2

\appendices

\section{Frame Efficiency Formulation with BNC}
\label[appendix]{sec:analysis}

For any real $c$ and $m \in \mathbb{Z}^+$, define $(c)^+ := \max\{c, 0\}$, 
$[m] := \{0, 1, \ldots, m\}$, and $\llbracket m \rrbracket := \{1, 2, \ldots, m\}$.
For any subset $Z$ of integers, define $mZ := \{mz \colon z \in Z\}$.
To formulate the frame efficiency with BNC, we consider $M > 1$.
Let $g = \gcd(M, N)$.
We separate the analysis into two cases.

\subsection{Case I: $M \le N$}

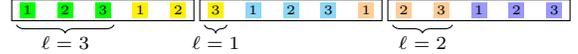
\begin{figure}
    \centering
    \scriptsize
    \begin{tikzpicture} 
        \foreach \j in {1,...,3}
        {
            \fill[green] (0.5*\j-0.5,0) rectangle node[name=a\j,color=black] {{\tiny $\j$}} ++(6pt,6pt);
        }
        \foreach \j in {1,...,3}
        {
            \fill[yellow] (1.5+0.5*\j-0.5,0) rectangle node[name=b\j,color=black] {{\tiny $\j$}} ++(6pt,6pt);
        }
        \foreach \j in {1,...,3}
        {
            \fill[cyan!40] (3+0.5*\j-0.5,0) rectangle node[name=c\j,color=black] {{\tiny $\j$}} ++(6pt,6pt);
        }
        \foreach \j in {1,...,3}
        {
            \fill[orange!40] (4.5+0.5*\j-0.5,0) rectangle node[name=d\j,color=black] {{\tiny $\j$}} ++(6pt,6pt);
        }
        \foreach \j in {1,...,3}
        {
            \fill[blue!40] (6+0.5*\j-0.5,0) rectangle node[name=e\j,color=black] {{\tiny $\j$}} ++(6pt,6pt);
        }
        \draw ($(a1)+(-6pt,4pt)$) rectangle node {} ($(b2)+(6pt,-4pt)$);
        \draw ($(b3)+(-6pt,4pt)$) rectangle node {} ($(d1)+(6pt,-4pt)$);
        \draw ($(d2)+(-6pt,4pt)$) rectangle node {} ($(e3)+(6pt,-4pt)$);
        \draw [decorate,decoration={brace,amplitude=3pt,mirror}] ($(a1.south west)+(0,-.05)$) -- ($(a3.south east)+(0,-.05)$) node [black,midway,below=1pt,minimum height=4mm] {$\ell = 3$};
        \draw [decorate,decoration={brace,amplitude=3pt,mirror}] ($(b3.south west)+(0,-.05)$) -- ($(b3.south east)+(0,-.05)$) node [black,midway,below=1pt,minimum height=4mm] {$\ell = 1$};
        \draw [decorate,decoration={brace,amplitude=3pt,mirror}] ($(d2.south west)+(0,-.05)$) -- ($(d3.south east)+(0,-.05)$) node [black,midway,below=1pt,minimum height=4mm] {$\ell = 2$};
    \end{tikzpicture}
    \vskip -1em
    \caption{An illustration of $\ell$ when $M = 3$ and $N = 5$. Each colored square is a BNC packet. The BNC packets with the same color belong to the same batch. Each rectangle represents the BNC packets that are aggregated in the same UDP-Lite packet.}
    \label{fig:case1}
\end{figure}

\begin{figure}
    \centering
    \scriptsize
    \begin{tikzpicture} 
        \foreach \j in {1,...,2}
        {
            \fill[green] (0.5*\j-0.5,0) rectangle node[name=a\j,color=black] {{\tiny $\j$}} ++(6pt,6pt);
        }
        \foreach \j in {1,...,2}
        {
            \fill[yellow] (1+0.5*\j-0.5,0) rectangle node[name=b\j,color=black] {{\tiny $\j$}} ++(6pt,6pt);
        }
        \foreach \j in {1,...,2}
        {
            \fill[cyan!40] (2+0.5*\j-0.5,0) rectangle node[name=c\j,color=black] {{\tiny $\j$}} ++(6pt,6pt);
        }
        \foreach \j in {1,...,2}
        {
            \fill[orange!40] (3+0.5*\j-0.5,0) rectangle node[name=d\j,color=black] {{\tiny $\j$}} ++(6pt,6pt);
        }
        \foreach \j in {1,...,2}
        {
            \fill[blue!40] (4+0.5*\j-0.5,0) rectangle node[name=e\j,color=black] {{\tiny $\j$}} ++(6pt,6pt);
        }
        \draw ($(a1)+(-6pt,4pt)$) rectangle node {} ($(c1)+(6pt,-4pt)$);
        \draw ($(c2)+(-6pt,4pt)$) rectangle node {} ($(e2)+(6pt,-4pt)$);
        \draw [decorate,decoration={brace,amplitude=3pt,mirror}] ($(a1.south west)+(0,-.05)$) -- ($(a2.south east)+(0,-.05)$) node [black,midway,below=1pt,minimum height=4mm] {$\ell = 2$};
        \draw [decorate,decoration={brace,amplitude=3pt,mirror}] ($(c2.south west)+(0,-.05)$) -- ($(c2.south east)+(0,-.05)$) node [black,midway,below=1pt,minimum height=4mm] {$\ell = 1$};
    \end{tikzpicture}
    \vskip -1em
    \caption{An illustration of packing $\left\lfloor \frac{N-\ell}{M} \right\rfloor$ complete batches in a UDP-Lite packet, where $M = 2$ and $N = 5$.}
    \label{fig:case1a}
\end{figure}
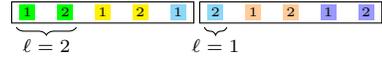

\begin{figure*}
    \centering
    \scriptsize
    \begin{tikzpicture} 
        \foreach \j in {1,...,7}
        {
            \fill[green] (0.5*\j-0.5,0) rectangle node[name=a\j,color=black] {{\tiny $\j$}} ++(6pt,6pt);
        }
        \foreach \j in {1,...,7}
        {
            \fill[yellow] (3.5+0.5*\j-0.5,0) rectangle node[name=b\j,color=black] {{\tiny $\j$}} ++(6pt,6pt);
        }
        \foreach \j in {1,...,7}
        {
            \fill[cyan!40] (7+0.5*\j-0.5,0) rectangle node[name=c\j,color=black] {{\tiny $\j$}} ++(6pt,6pt);
        }
        \foreach \j in {1,...,7}
        {
            \fill[orange!40] (10.5+0.5*\j-0.5,0) rectangle node[name=d\j,color=black] {{\tiny $\j$}} ++(6pt,6pt);
        }
        \foreach \j in {1,...,7}
        {
            \fill[blue!40] (14+0.5*\j-0.5,0) rectangle node[name=e\j,color=black] {{\tiny $\j$}} ++(6pt,6pt);
        }
        \draw ($(a1)+(-6pt,4pt)$) rectangle node {} ($(a5)+(6pt,-4pt)$);
        \draw ($(a6)+(-6pt,4pt)$) rectangle node {} ($(b3)+(6pt,-4pt)$);
        \draw ($(b4)+(-6pt,4pt)$) rectangle node {} ($(c1)+(6pt,-4pt)$);
        \draw ($(c2)+(-6pt,4pt)$) rectangle node {} ($(c6)+(6pt,-4pt)$);
        \draw ($(c7)+(-6pt,4pt)$) rectangle node {} ($(d4)+(6pt,-4pt)$);
        \draw ($(d5)+(-6pt,4pt)$) rectangle node {} ($(e2)+(6pt,-4pt)$);
        \draw ($(e3)+(-6pt,4pt)$) rectangle node {} ($(e7)+(6pt,-4pt)$);
        \draw [decorate,decoration={brace,amplitude=3pt,mirror}] ($(a1.south west)+(0,-.05)$) -- ($(a5.south east)+(0,-.05)$) node [black,midway,below=1pt,minimum height=4mm] {$\ell = 5$};
        \draw [decorate,decoration={brace,amplitude=3pt,mirror}] ($(a6.south west)+(0,-.05)$) -- ($(a7.south east)+(0,-.05)$) node [black,midway,below=1pt,minimum height=4mm] {$\ell = 2$};
        \draw [decorate,decoration={brace,amplitude=3pt,mirror}] ($(b4.south west)+(0,-.05)$) -- ($(b7.south east)+(0,-.05)$) node [black,midway,below=1pt,minimum height=4mm] {$\ell = 4$};
        \draw [decorate,decoration={brace,amplitude=3pt,mirror}] ($(c2.south west)+(0,-.05)$) -- ($(c6.south east)+(0,-.05)$) node [black,midway,below=1pt,minimum height=4mm] {$\ell = 5$};
        \draw [decorate,decoration={brace,amplitude=3pt,mirror}] ($(c7.south west)+(0,-.05)$) -- ($(c7.south east)+(0,-.05)$) node [black,midway,below=1pt,minimum height=4mm] {$\ell = 1$};
        \draw [decorate,decoration={brace,amplitude=3pt,mirror}] ($(d5.south west)+(0,-.05)$) -- ($(d7.south east)+(0,-.05)$) node [black,midway,below=1pt,minimum height=4mm] {$\ell = 3$};
        \draw [decorate,decoration={brace,amplitude=3pt,mirror}] ($(e3.south west)+(0,-.05)$) -- ($(e7.south east)+(0,-.05)$) node [black,midway,below=1pt,minimum height=4mm] {$\ell = 5$};
        \draw [decorate,decoration={brace,amplitude=3pt}] ($(a1.north west)+(-.2,.05)$) -- ($(a1.north west)+(0,.05)$) node [black,midway,above=1pt,minimum height=4mm] {$s = 0$};
        \draw [decorate,decoration={brace,amplitude=3pt}] ($(b1.north west)+(0,.05)$) -- ($(b3.north east)+(0,.05)$) node [black,midway,above=1pt,minimum height=4mm] {$s = 3$};
        \draw [decorate,decoration={brace,amplitude=3pt}] ($(c1.north west)+(0,.05)$) -- ($(c1.north east)+(0,.05)$) node [black,midway,above=1pt,minimum height=4mm] {$s = 1$};
        \draw [decorate,decoration={brace,amplitude=3pt}] ($(d1.north west)+(0,.05)$) -- ($(d4.north east)+(0,.05)$) node [black,midway,above=1pt,minimum height=4mm] {$s = 4$};
        \draw [decorate,decoration={brace,amplitude=3pt}] ($(e1.north west)+(0,.05)$) -- ($(e2.north east)+(0,.05)$) node [black,midway,above=1pt,minimum height=4mm] {$s = 2$};
    \end{tikzpicture}
    \vskip -1em
    \caption{An illustration of $\ell$ and $s$ when $M = 7$ and $N = 5$. Except $\ell = N$, every $\ell \in g \left\llbracket \frac{N-g}{g} \right\rrbracket$ appears exactly once in a period of $\lcm(M, N)$ BNC packets.
    }
    \label{fig:case2}
\end{figure*}
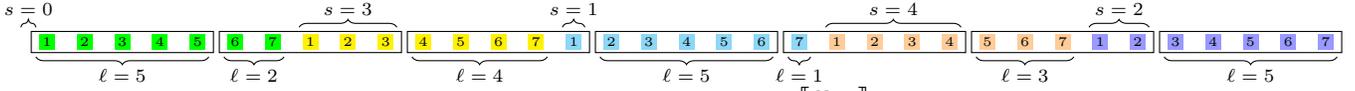

Let $\ell$ be the number of BNC packets of the first batch packed in this UDP-Lite packet.
An example of $\ell$ is illustrated in \Cref{fig:case1}.
The value $\ell$ is the remainder of some multiple of $N$ divided by $M$, but with the remainder $0$ mapped to $M$.
That is, there exist some integers $x, y$ such that $xN - yM = \ell$.
This is a linear Diophantine equation in two unknowns.
Its solution $(x, y)$ exists if and only if $\ell$ is a multiple of $g$.
Each $\ell \in g \left\llbracket \frac{M}{g} \right\rrbracket$ appears exactly once in a period of $\lcm(M, N)$ BNC packets.

Consider a new batch, i.e., the next node has not received any BNC packet of this batch yet.
Suppose $\ell$ BNC packets of this batch is packed in a UDP-Lite packet.
Let $\beta'(\ell)$ be the expected rank increment of this batch at the next node after the current node sends this UDP-Lite packet, given that the payload of this UDP-Lite packet is passed to the BNC recoder/decoder.
Then, we have
\begin{equation*}
    \beta'(\ell) = \sum_{r = 1}^M \hbar_r \sum_{j = 0}^\ell \bin(j; \ell, f) \min\{j, r\}.
\end{equation*}
On the other hand, consider a batch that $M-\ell$ BNC packets of this batch were packed and have been sent in an earlier UDP-Lite packet.
The remaining $\ell$ BNC packets of this batch is packed in a new UDP-Lite packet.
Let $\beta(\ell)$ be the expected rank increment of this batch at the next node after the current node sends this new UDP-Lite packet, given that the payload of this UDP-Lite packet is passed to the BNC recoder/decoder.
We have
\begin{gather*}
    \beta(\ell) = \sum_{r = 1}^M \hbar_r \sum_{j = 0}^\ell \bin(j; \ell, f) \smashoperator{\sum_{i = 0}^{M-\ell}} \bin_d(i; M-\ell, f) (\min\{j, r-i\})^+.
\end{gather*}

Consider a single UDP-Lite packet that consists of $N$ aggregated BNC packets.
These $N$ BNC packets can be from different batches.
First, $\ell$ BNC packets of the first batch are packed, where the remaining $M-\ell$ BNC packets of this batch were sent already.
If $N-\ell \ge M$, then we can pack some complete batches (all $M$ BNC packets) to this UDP-Lite packet.
We can totally pack $\left\lfloor \frac{N-\ell}{M} \right\rfloor$ complete batches.
The remaining $(N-\ell) \bmod M$ BNC packets in this UDP-Lite packet, if there is any, are from a new batch.
This is illustrated in \Cref{fig:case1a}.

Let $q_\ell$ be the probability of the occurrence of $\ell$,
i.e., $q_\ell = \frac{g}{M}$ for each $\ell \in g \left\llbracket \frac{M}{g} \right\rrbracket$.
The efficiency optimization can be formulated as 
\begin{multline*}
    \max_{N \in \mathbb{Z}^+} \frac{dK}{S(N)} \sum_{\ell \in g \left\llbracket \frac{M}{g} \right\rrbracket} q_\ell \left[ \beta(\ell) + \beta'((N-\ell) \bmod M) \vphantom{\left\lfloor \frac{N-\ell}{M} \right\rfloor} \right. \\
    \left. +\; \left\lfloor \frac{N-\ell}{M} \right\rfloor \beta'(M) \right] \quad \text{s.t.} \quad P+N(H+K+F) \le L.
\end{multline*}

After factoring out $q_\ell = \frac{g}{M}$ from the summation, the sum considers all batches within the period of $\lcm(M, N)$ BNC packets.
Due to the periodic structure, we have
\begin{equation*}
    \sum_{\ell \in g \left\llbracket \frac{M}{g} \right\rrbracket} \left\lfloor \frac{N-\ell}{M} \right\rfloor = \frac{\lcm(M, N) - \frac{M-g}{g}M}{M} = \frac{N-M+g}{g}.
\end{equation*}
The optimization problem becomes
\begin{IEEEeqnarray*}{Cl}
    \max_{N \in \mathbb{Z}^+} & \frac{dK}{MS(N)} \left( (N-M+g) \beta'(M) + \smashoperator{\sum_{\ell \in g \left\llbracket \frac{M}{g} \right\rrbracket}} g(\beta(\ell) + \beta'(\ell)) \right)\\
    \text{s.t.} & P+N(H+K+F) \le L.
\end{IEEEeqnarray*}

\subsection{Case II: $M > N$}

Similar to Case I, let $\ell$ be the number of BNC packets of the first batch packed in this UDP-Lite packet.
The value of $\ell$ is either $N$ or the remainder of some multiple of $M$ divided by $N$ excluding $0$.
Again, $\ell$ must be a multiple of $g$.
Each $\ell \in g \left\llbracket \frac{N-g}{g} \right\rrbracket$ appears once in the period of $\lcm(M, N)$ BNC packets.
However, $\ell = N$ may appear many times.
\Cref{fig:case2} illustrates an example of $\ell$.

When $\ell \in g \left\llbracket \frac{N-g}{g} \right\rrbracket$, there can only be one more batch in the same UDP-Lite packet.
This batch has $N-\ell$ BNC packets packed in this UDP-Lite packet.
Therefore, the number of $\ell = N$ within a period of $\lcm(M, N)$ BNC packets is $\frac{\lcm(M, N)-\frac{N-g}{g}N}{N} = \frac{M-N+g}{g}$.
Let $q_\ell$ be the probability of the occurrence of $\ell$.
As there are totally $\frac{\lcm(M, N)}{N} = \frac{M}{g}$ UDP-Lite packets in a period, we have
\begin{equation*}
    q_\ell = \begin{cases}
        \frac{g}{M} & \text{if } \ell \in g \left\llbracket \frac{N-g}{g} \right\rrbracket,\\
        \frac{M-N+g}{M} & \text{if } \ell = N.
    \end{cases}
\end{equation*}

Consider a batch that $M-\ell$ BNC packets of this batch have been sent in earlier $\left\lceil \frac{M-\ell}{N} \right\rceil$ UDP-Lite packets.
The remaining $\ell$ BNC packets of this batch are aggregated in a new UDP-Lite packet.
Let $\gamma(\ell)$ be the expected rank increment of this batch at the next node after the current node sends this new UDP-Lite packet, given that the payload of this UDP-Lite packet is passed to the BNC recoder/decoder.
We have
\begin{gather*}
    \gamma(\ell) = \sum_{r = 1}^M \hbar_r \sum_{j = 0}^\ell \bin(j; \ell, f) \omega_{r, j, (M-\ell) \bmod N}\left(\left\lfloor \frac{M-\ell}{N} \right\rfloor, 0\right),
\end{gather*}
where
$\omega_{r, j, \bar{b}}(\alpha, \mu)$
\begin{equation*}
    = \begin{cases}
        \sum_{i = 0}^N \bin_d(i; N, f) \omega_{r, j, \bar{b}}(\alpha - 1, \mu+i) & \text{if } \alpha > 0,\\
        \sum_{i = 0}^{\bar{b}} \bin_d(i; \bar{b}, f) (\min\{j, r - (\mu+i)\})^+ & \text{if } \alpha = 0.
    \end{cases}
\end{equation*}

The recursive relation $\omega_{r, j, (M-\ell) \bmod N}\left(\left\lfloor \frac{M-\ell}{N} \right\rfloor, 0\right)$ can be interpreted as the expected rank increment of a batch at the next node by receiving $j$ out of $\ell$ BNC packets of this batch, 
where the remaining BNC packets of this batch has been sent in earlier $\left\lceil \frac{M-\ell}{N} \right\rceil$ UDP-Lite packets.
In these earlier UDP-Lite packets, this batch fully occupies $\left\lfloor \frac{M-\ell}{N} \right\rfloor$ of them, and partially packed $(M-\ell) \bmod N$ BNC packets (if $>0$) in the earliest one.
Each layer of recursion corresponds to one of the $\left\lceil \frac{M-\ell}{N} \right\rceil$ already sent UDP-Lite packets

\begin{figure}
    \centering
    \scriptsize
    \begin{tikzpicture}[xscale=.7] 
        \foreach \j in {1,...,8}
        {
            \fill[green] (0.5*\j-0.5,0) rectangle node[name=a\j,color=black] {{\tiny $\j$}} ++(6pt,6pt);
        }
        \foreach \j in {1,...,8}
        {
            \fill[yellow] (4+0.5*\j-0.5,0) rectangle node[name=b\j,color=black] {{\tiny $\j$}} ++(6pt,6pt);
        }
        \foreach \j in {1,...,8}
        {
            \fill[cyan!40] (8+0.5*\j-0.5,0) rectangle node[name=c\j,color=black] {{\tiny $\j$}} ++(6pt,6pt);
        }
        \draw ($(a1)+(-6pt,4pt)$) rectangle node {} ($(a6)+(6pt,-4pt)$);
        \draw ($(a7)+(-6pt,4pt)$) rectangle node {} ($(b4)+(6pt,-4pt)$);
        \draw ($(b5)+(-6pt,4pt)$) rectangle node {} ($(c2)+(6pt,-4pt)$);
        \draw ($(c3)+(-6pt,4pt)$) rectangle node {} ($(c8)+(6pt,-4pt)$);
        \draw [decorate,decoration={brace,amplitude=3pt,mirror}] ($(a1.south west)+(0,-.05)$) -- ($(a6.south east)+(0,-.05)$) node [black,midway,below=1pt,minimum height=4mm] {$\ell = 6$};
        \draw [decorate,decoration={brace,amplitude=3pt,mirror}] ($(a7.south west)+(0,-.05)$) -- ($(a8.south east)+(0,-.05)$) node [black,midway,below=1pt,minimum height=4mm] {$\ell = 2$};
        \draw [decorate,decoration={brace,amplitude=3pt,mirror}] ($(b5.south west)+(0,-.05)$) -- ($(b8.south east)+(0,-.05)$) node [black,midway,below=1pt,minimum height=4mm] {$\ell = 4$};
        \draw [decorate,decoration={brace,amplitude=3pt,mirror}] ($(c3.south west)+(0,-.05)$) -- ($(c8.south east)+(0,-.05)$) node [black,midway,below=1pt,minimum height=4mm] {$\ell = 6$};
        \draw [decorate,decoration={brace,amplitude=3pt}] ($(a1.north west)+(-.2,.05)$) -- ($(a1.north west)+(0,.05)$) node [black,midway,above=1pt,minimum height=4mm] {$s = 0$};
        \draw [decorate,decoration={brace,amplitude=3pt}] ($(c1.north west)+(0,.05)$) -- ($(c2.north east)+(0,.05)$) node [black,midway,above=1pt,minimum height=4mm] {$s = 2$};
    \end{tikzpicture}
    \vskip -1em
    \caption{An example showing that the range of $s$ is not always $g \left[ \frac{N-g}{g} \right]$, where $M = 8$ and $N = 6$. The second (yellow) batch does not have an $s = 4$ because it does not give an $\ell = N$ in any UDP-Lite packet.}
    \label{fig:case2a}
\end{figure}
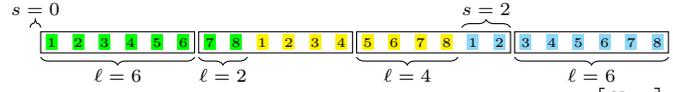

Now, consider that the current UDP-Lite packet only consists of a single batch, i.e., $\ell = N$.
There are two possibilities.
Some BNC packets of this batch may be packed in earlier UDP-Lite packets, or this batch first appears in this UDP-Lite packet.
Suppose this batch has $s'$ BNC packets packed in the UDP-Lite packet that this batch first appears.
Let $s = s' \mod N$, i.e., we set $s = 0$ when $s' = N$.
Excluding this UDP-Lite packet, the remaining number of BNC packets for this batch is $M-s$.
    If $M-s < N$, this invalidates the assumption $\ell = N$.
    So, when $\ell = N$, the valid range of $s$ is not always $g \left[ \frac{N-g}{g} \right]$.
    An example is illustrated in \Cref{fig:case2a}.
The valid range of $s$ is $g \left[ \frac{\min\{M-N, N-g\}}{g} \right]$ instead.

For each $s$, let $k$ be the number of earlier UDP-Lite packets, each of which has $N$ packets of this batch aggregated.
The valid range of each $k$ given $s$ is $\left[ \left\lfloor \frac{M-s}{N} \right\rfloor - 1 \right]$.
For simplicity, let $\mathbb{S}$ be the set of all valid $(s, k)$ pairs.
Each valid $(s, k) \in \mathbb{S}$ appears once in each period of $\lcm(M, N)$ BNC packets.
In other words, the probability of appearing a valid pair of $(s, k)$ is $q_{s, k} = \frac{g}{M}$.

Given $s, k$ and $\ell = N$, let $\gamma'(s, k)$ be the expected rank increment at the next node after the current node sends this UDP-Lite packet, given that the payload of this UDP-Lite packet is passed to the BNC recoder/decoder.
We have
\begin{equation*}
    \gamma'(s, k) = \sum_{r = 1}^M \hbar_r \sum_{j = 0}^N \bin(j; N, f) \omega_{r, j, s}(k, 0).
\end{equation*}
Therefore, the efficiency optimization is 
\begin{IEEEeqnarray*}{Cl}
    \max_{N \in \mathbb{Z}^+} & \frac{dK}{S(N)} \left( \sum_{\ell \in g \left\llbracket \frac{N-g}{g} \right\rrbracket} \!\!\!\!\!\! q_\ell (\gamma(\ell) + \beta'(N-\ell)) + \smashoperator{\sum_{(s, k) \in \mathbb{S}}} q_{s, k} \gamma'(s, k) \right)\\
    \text{s.t.} & P+N(H+K+F) \le L,
\end{IEEEeqnarray*}
which can be simplified as
\begin{IEEEeqnarray*}{Cl}
    \max_{N \in \mathbb{Z}^+} & \frac{dKg}{MS(N)} \left( \sum_{\ell \in g \left\llbracket \frac{N-g}{g} \right\rrbracket} \!\!\!\!\!\! (\gamma(\ell)+\beta'(N-\ell)) + \smashoperator{\sum_{(s, k) \in \mathbb{S}}} \gamma'(s, k) \right)\\
    \text{s.t. } & P+N(H+K+F) \le L.
\end{IEEEeqnarray*}

\fi

\bibliographystyle{IEEEtran}
\bibliography{ref}

\end{document}